\documentclass{ragtime} 

\usepackage{amsfonts}

\usepackage{times}

\title[Equilibrium tori orbiting Reissner-Nordström naked singularities]%
      {Equilibrium tori orbiting Reissner-Nordström naked singularities}

\author[R. Mishra, % Run. head authors: separate names with commas,
        and W\l odek Klu\'{z}niak]%    Now let's start the paper title authors:
       {Ruchi Mishra\at{a} % Makes referencing superscript `1'
         W\l odek Klu\'{z}niak\at[]{}\\% Termination of authors' block; if
        \ins{}Nicolaus Copernicus Astronomical Center, Polish academy of Sciences,\\
        ~ul. Bartycka 18, 00-716
Warsaw, Poland\splitins[1]% This is how to break an
       \\% Termination of the first affiliation.
        \\
        \ins{a}\Email{rmishra@camk.edu.pl}}
\coentry{Z. Stuchl\'{\i}k, J. Kov\'a\v{r} and J. Cimrman}%
       {Equilibrium of spinning test particles in equatorial plane\par
        of Kerr--de~Sitter spacetimes}

\begin{document}

% Citation of references in abstract should generally be avoided to
% ensure self-consistency of the abstract.  If you do insist on citation(s)
% within the abstract, you should use the \bibentry command, which forces
% the _complete_bibliographic_entry_ to appear in the abstract.
% With the `nonatbib' optional class argument this feature is not available.
\begin{abstract}
In general relativity, the asymptotically flat space-time of a charged, spherically symmetric (non-rotating) body is described by the Reissner-Nordström metric. This metric corresponds to a naked singularity when the absolute value of charge, $Q$, exceeds the mass, $M$. For all Reissner-Nordström naked singularities, there exists a zero gravity sphere where a test particle can remain at rest. Outside that sphere gravity is attractive, inside it gravity is repulsive. For values of $Q/M>\sqrt{9/8}$ the angular frequency of circular test-particle orbits has a maximum at radius $r=(4/3)\,Q^2/M$. We construct polytropic tori with uniform values of specific angular momentum in the naked singularity regime of the Reissner-Nordström metric, $(Q/M>1)$.
\end{abstract}

% The key words are to be separated by the N-dash surrounded by spaces,
% the left one non-breakable.  The name concatenation like Kerr--de~Sitter
% should also be typed with N-dash but with no spaces around (compare,
% e.g., Levi-Civita, which is a single person):

% It is good to provide as many as \label's possible, but never start the
% key with a numeral.  This makes problems with pdflatex processing.
\section*{Introduction}\label{intro}%%%%%%%%%%%%%%%%%%%%%%%%%%%%%%%%%%%%%%%%
The gravity of a spherically symmetric, electrically charged body is described by the  Reissner–Nordström (RN) space-time metric. For values of charge $Q>M$ the metric describes a naked singularity. One of the key features of many spherically symmetric naked singularities is the existence of a ``zero-gravity'' spherical surface, inside of which gravity is repulsive, whereas outside of it gravity is attractive as usual \citep{2011Remo, 2014KS, 2023VK}.  This surface marks the innermost boundary for the existence of circular geodesics, representing orbits with zero angular momentum, a test particle remaining stably at rest there. In the RN case, no circular photon orbits exist for $Q/M>\sqrt{9/8}$. For a somewhat larger value still of $Q/M$ no marginally stable orbits exist, so test-particle circular orbits are stable all the way down to the zero-gravity sphere (Figure~1).

In a standard thin accretion disk, accretion is enabled by (effective) viscous torques transferring angular momentum outwards. This is possible as long as angular frequency decreases monotonically with the radius. In the case of RN space-time, the torque is reversed close to the zero-gravity radius, $r_0$, as $d\Omega/dr>0$ there, for test particles. It is not yet clear whether a thin disk may exist all the way to the zero-gravity sphere. For this reason we consider, instead, toroidal orbiting configurations of a perfect fluid.

%When we approach the maximum of angular frequency (at the marginally stable orbit), $r_{\Omega\mathrm{max}}$, where Keplerian accretion ceases to be effective, material is likely to accumulate, giving rise to formation of a fluid toroidal structure. 
%The equilibrium configurations of toroidal perfect fluid can be connected to the inner region of circular geodesics. In this chapter 

Here, we focus on the structure and shape\footnote{While the configurations have the topology of a torus, their cross-section is not circular, not even ellipsoidal.} of equilibrium tori in which a perfect fluid is orbiting the RN naked singularity. As a first step, we consider uniform angular momentum distributions. Earlier work by \cite{2015Sk,2011JKZ,2023P} presented the equilibrium toroidal structures in the case of Kehagias–Sfetsos naked singularities, Reissner-Nordström-(anti-)de Sitter space times, and q-metric space time respectively. Our results are qualitatively similar to \citep{2015Sk}. We use units with $G=1, c=1$. For convenience, since our results depend only on $(Q/M)^2$ we take $Q>0$ to be the absolute magnitude of charge, the actual charge being $\pm Q$.
%############################################################################################################################################################################################################################################
\begin{figure}
    \centering
    \includegraphics[width=1\columnwidth,height=0.65\columnwidth]{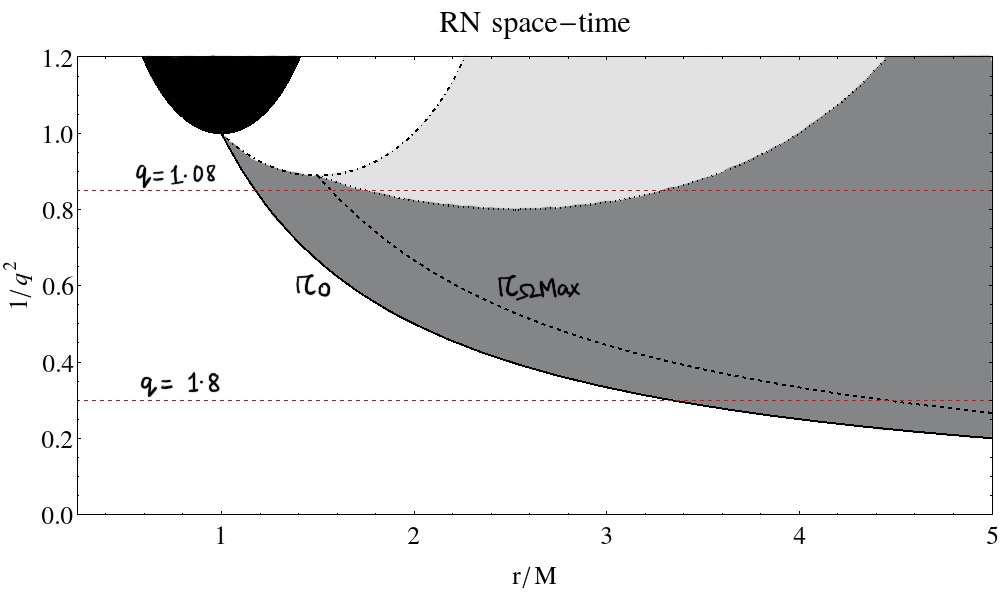}
    \caption{Orbital stability diagram  for the RN spacetime ($q\equiv Q/M$). The {\sl black} region corresponds to the hypervolume between the event horizons. The {\sl solid black} line represents the zero gravity sphere (radius $r_0$). The {\sl black dashed } lines represents the radius, $r_{\Omega\mathrm{max}}$, at which angular frequency is maximum. The black {\sl dotted} line corresponds to the marginally stable orbit whereas the {\sl dot-dashed} line corresponds to the photon orbit. {\sl Dark grey} region corresponds to the stability region for time-like circular geodesics. {\sl Light grey} region corresponds to the instability region for time-like circular geodesics. No circular geodesics exist in the {\sl white} regions. The horizontal {\sl red dashed} lines correspond to the two choices of $Q/M$ ratio discussed in this contribution.
    }
    \label{stability}
\end{figure}
%############################################################################################################################################################################################################################################
%%%%%%%%%%%%%%%%%%%%%%%%%%%%%%%%%%%%%%%%%%%%%%%%
\section*{Geodesic motion in RN space time}
%%%%%%%%%%%%%%%%%%%%%%%%%%%%%%%%%%%%%%%%%%%%%%%%
 The RN metric can be expressed in standard spherical coordinates as 
 \begin{equation}
     ds^2 = -f(r)\,dt^2 + f^{-1}(r)\,dr^2+r^2(d\theta^2+ \sin^2\theta \,d\phi^2),
 \end{equation}
 where $f(r)$ is the metric function given by 
 \begin{equation}
 f(r) = 1 - \frac{2M}{r} +\frac{Q^2}{r^2} .
 \end{equation}

%############################################################################################################################################################################################################################################
\begin{figure}
    \centering
    \includegraphics[width=1\columnwidth,height=0.65\columnwidth]{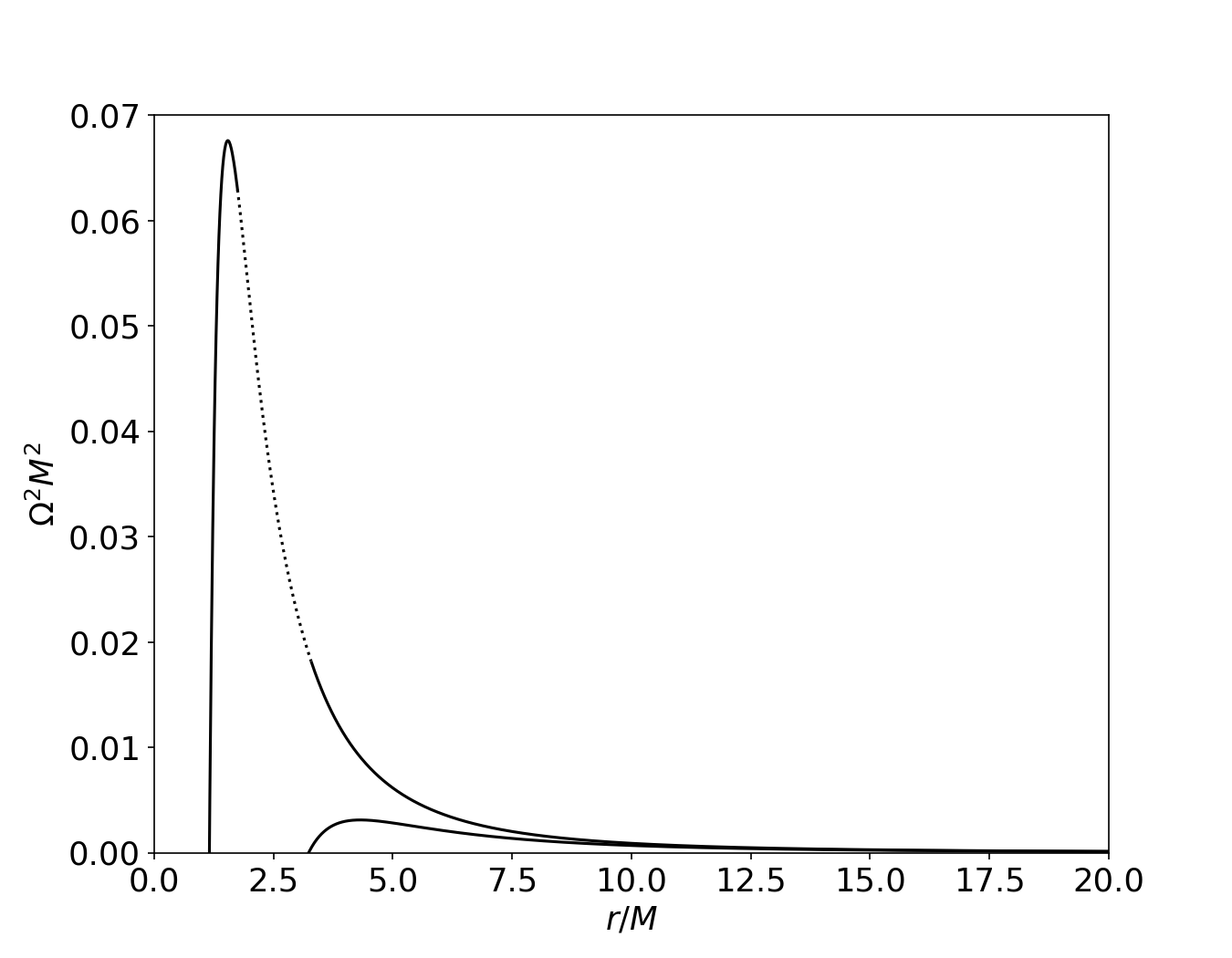}
            \caption{Square of orbital frequency as a function of $r/M$ for charge value $Q/M = 1.08$ {\sl (top)} and $Q/M=1.8$ {\sl (bottom)}. The dotted line represents the unstable region.}
    \label{omega}
\end{figure}
%############################################################################################################################################################################################################################################

%############################################################################################################################################################################################################################################
\begin{figure}
    \centering
    \includegraphics[width=1\columnwidth,height=0.65\columnwidth]{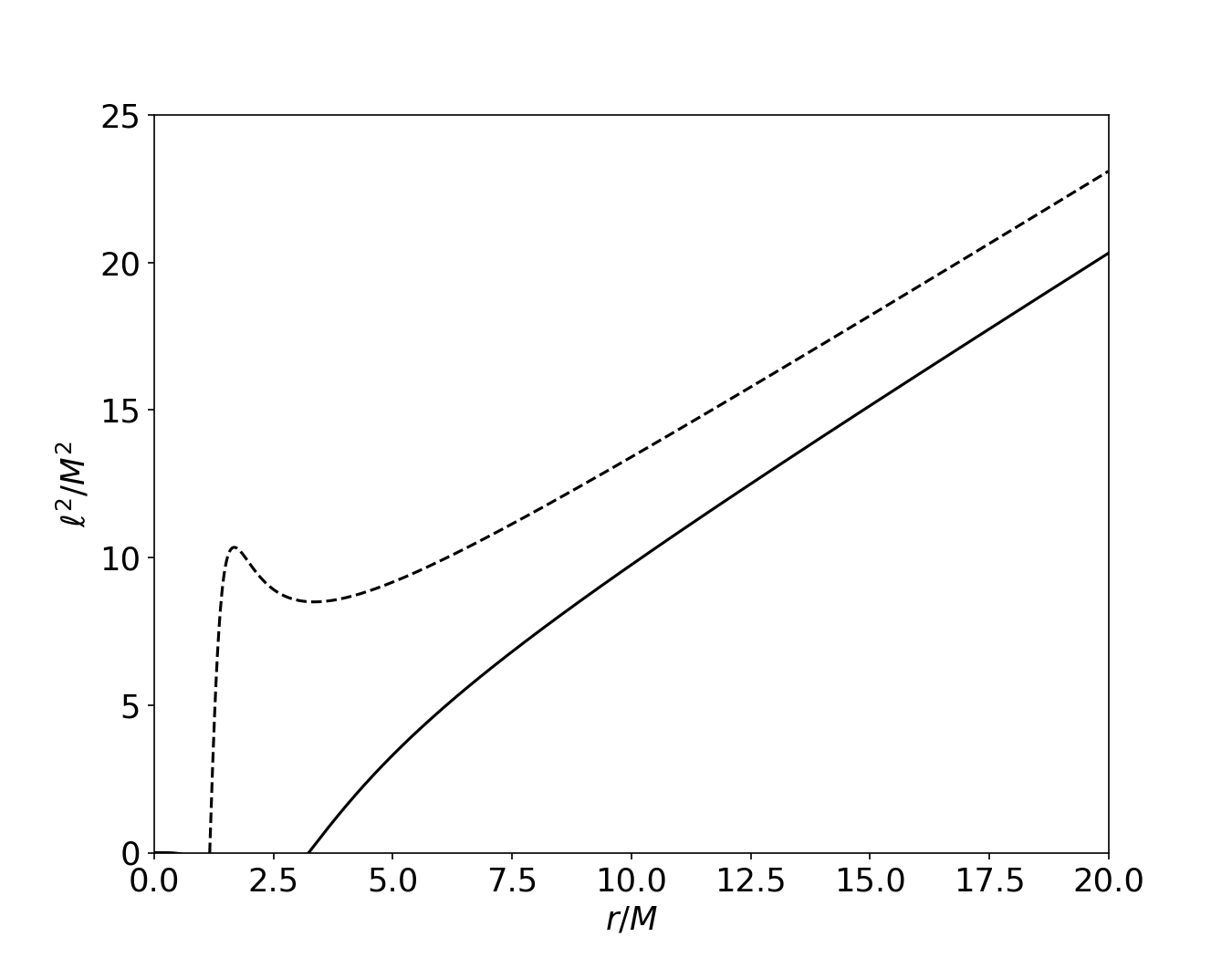}
    \caption{Square of specific angular momentum as a function of $r/M$ for charge value $Q/M = 1.08$ (dashed) and $Q/M=1.8$ (solid).}
    \label{omega}
\end{figure}
%############################################################################################################################################################################################################################################
 For a perfect fluid, the energy momentum tensor is given by

 \begin{equation}
     T_{\mu\nu} = (p+\rho)u_\mu u_\nu -pg_{\mu\nu},
 \end{equation}
where $\rho$ is the fluid energy density and $p$ its pressure. The four velocity of the fluid in circular orbit is $u^\mu =  (u^t,0,0,u^\phi)$.
The conserved energy and angular momentum of a test particle in this orbit are given by
\begin{equation}
    E \equiv -g_{tt}\dot{t} = -u_t,\\
%\end{equation}
%\begin{equation}
     L \equiv  g_{\phi\phi}\dot{\phi} = u_\phi.
\end{equation}
The conserved specific angular momentum $l$ is defined by
\begin{equation}
    l \equiv \frac{L}{E} = -\frac{u_\phi}{u_t}.
\end{equation}

We begin with a discussion of test particle circular motion. The effective potential can be conveniently taken to be $E^2$, in the equatorial plane ($\sin\theta=1$) it reduces to
\begin{equation}
    V_{\mathrm{eff}}(r)= f(r) \bigg(1 + \frac{L^2}{r^2}\bigg),
\end{equation}
where we've used $u_\mu u^\mu=-1$.
The condition $\partial  V_{\mathrm{eff}}/ \partial r = 0$  then gives  the angular momentum of circular geodesics, i.e.

\begin{equation}
    L^2(r) = \frac{r^3f'(r)}{2f-rf'(r)},
\end{equation}

or, 
\begin{equation}
    l^2 (r)= \frac{r^3}{2f^2} f'(r).
\end{equation}
 The orbital angular frequency, $\Omega=u^\phi/u^t$, so
\begin{equation}
    \Omega ^2 (r)= \frac{f'(r)}{2r}.
\end{equation}
The angular velocity for two values of $Q/M$ is illustrated in Fig.~2.

 For the RN space time the maximum of angular velocity is given by the condition
\begin{equation}
   \frac{d \Omega^2}{dr^2} = \frac{d}{dr}\bigg(\frac{f'(r)}{2r}\bigg) = 0,
\end{equation}
so the maximum of angular velocity occurs at
\begin{equation}
    r_{\Omega\mathrm{max}} = \frac{4}{3} \frac{Q^2}{M}.
\end{equation}
In Figure~1  we present this radius,  $r_{\Omega\mathrm{max}}$ as a function of $Q/M$. As circular geodesics are not possible within the photon orbit, $r_{\Omega\mathrm{max}}$ is only defined as long as $Q/M>\sqrt{9/8}$. Note that the marginally orbits, when they exist, are always at a larger radius---the maximum of $\Omega$
is always attained in a stable orbit.
The specific angular momentum at this orbit (of maximum $\Omega$) is given by,
\begin{align}
    l_\mathrm{\Omega max} \equiv l( r_{\Omega\mathrm{max}}).
\end{align}
It is interesting to note that the specific angular momentum and its radial derivative in circular orbits are not necessarily monotonic as a function of the radius $r$. The extrema of the $l(r)$ curve (if present) correspond to the marginally stable orbits. The region between the maximum and minimum of the $Q/M=1.08$ specific angular momentum curve in Figure~3 correspond to unstable circular orbits (c.f. Figure~2) by Rayleigh's criterion, as $dl/dr<0$  there.

In the case of barotropic fluid, the surfaces of constant pressure (isobars) corresponding to equipotential surfaces  $W(r, \theta)$ are given by Boyer’s condition
\citep[e.g.,][]{1978Ab,1978Koz} 
\begin{equation}
    -\int_{0}^{p} \frac{dp}{p+\rho} = W-W_{\mathrm{in}} = \ln\frac{u_t~}{(u_t)_{\mathrm{in}}} - \int_{l_\mathrm{in}}^{l}\frac{\Omega dl}{1-\Omega l}\,.
\end{equation}
In the Newtonian limit $W$ is the usual effective potential. Here  the subscript “in” refers to the inner edge of the torus. The equipotential surfaces are determined by the condition

\begin{equation}
    W(r,\theta) = \mathrm{constant}.
\end{equation}
%############################################################################################################################################################################################################################################
\begin{figure}
    \centering
    \includegraphics[width=1\columnwidth,height=0.65\columnwidth]{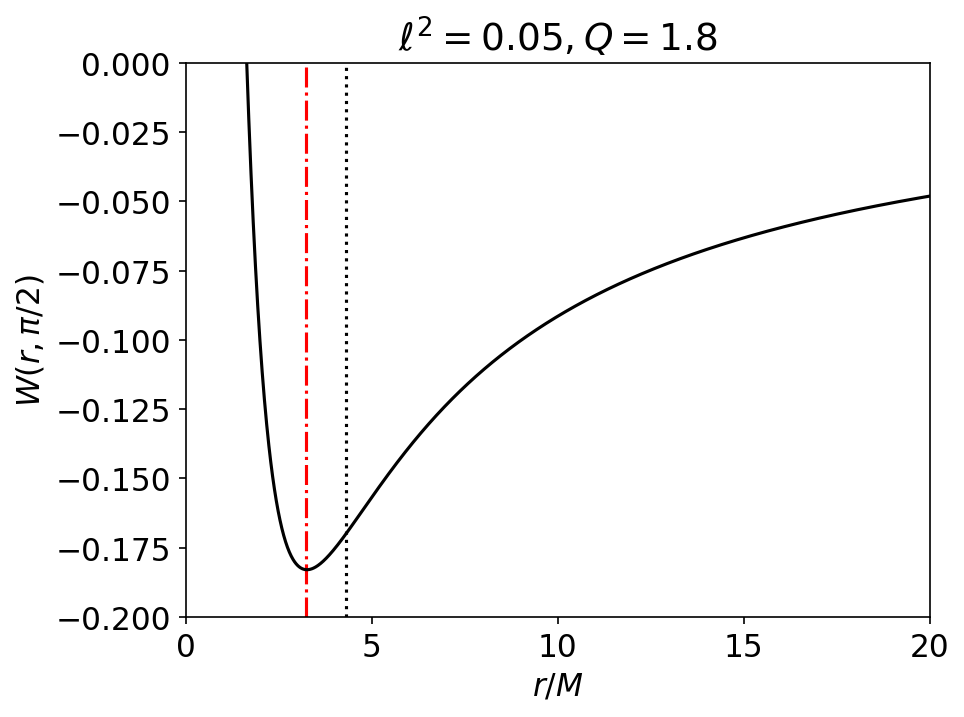}
    \includegraphics[width=1\columnwidth,height=0.65\columnwidth]{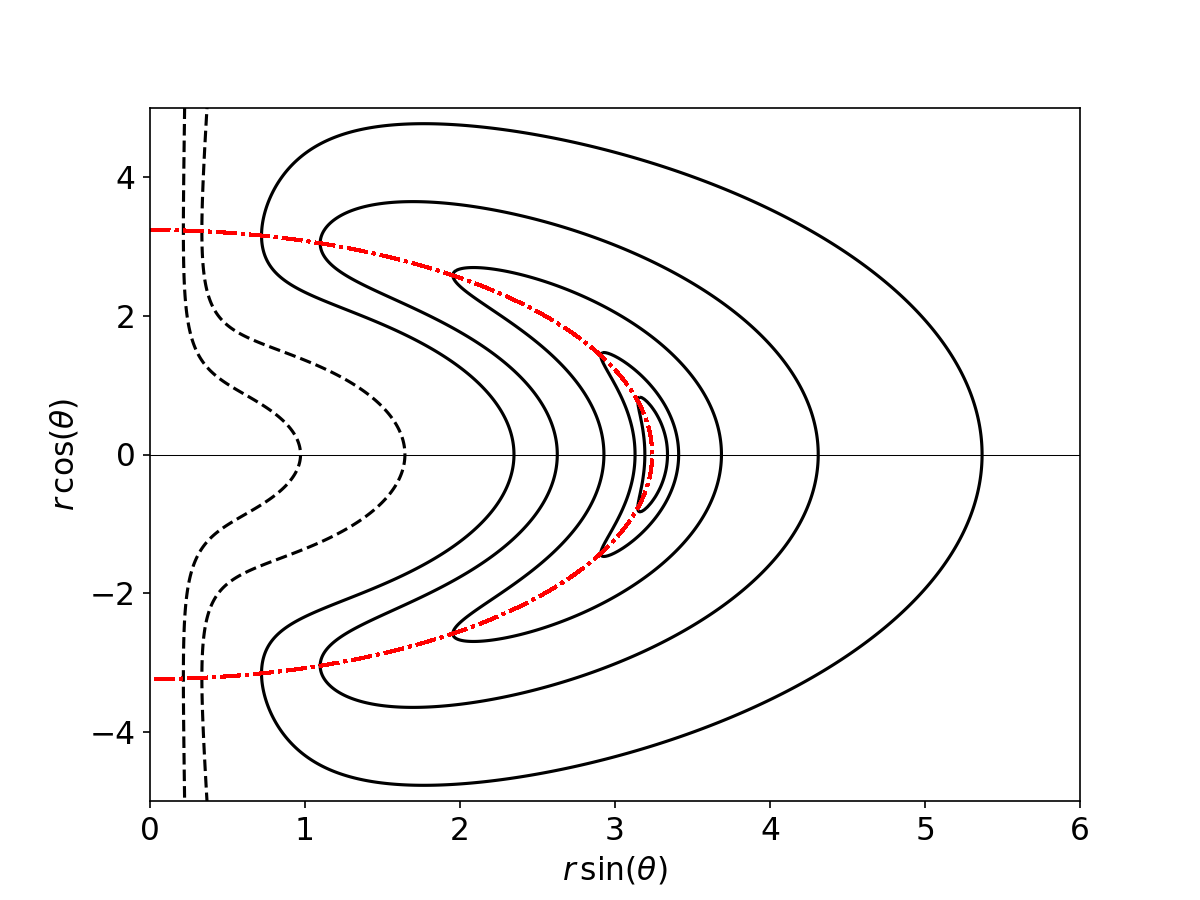}
    \caption{Equilibrium tori around RN naked singularity with $Q/M$=1.8 with $l_0 <l_\mathrm{\Omega max}$.  Location of the zero-gravity sphere is indicated by red dotted-dashed line. {\sl Top:} $W$ as a function of $r/M$ in equatorial plane. The radius of maximum angular frequency, $ r_{\Omega\mathrm{max}}$, is indicated by the black dotted line. {\sl Bottom:} Meridional cross-section of the equipotential surfaces.  Solid closed lines represent toroidal surfaces with negative $W$ value. The black dashed lines represent positive values of W.}
    \label{1a}
\end{figure}
%############################################################################################################################################################################################################################################

%############################################################################################################################################################################################################################################
\begin{figure}
    \centering
    \includegraphics[width=1\columnwidth,height=0.65\columnwidth]{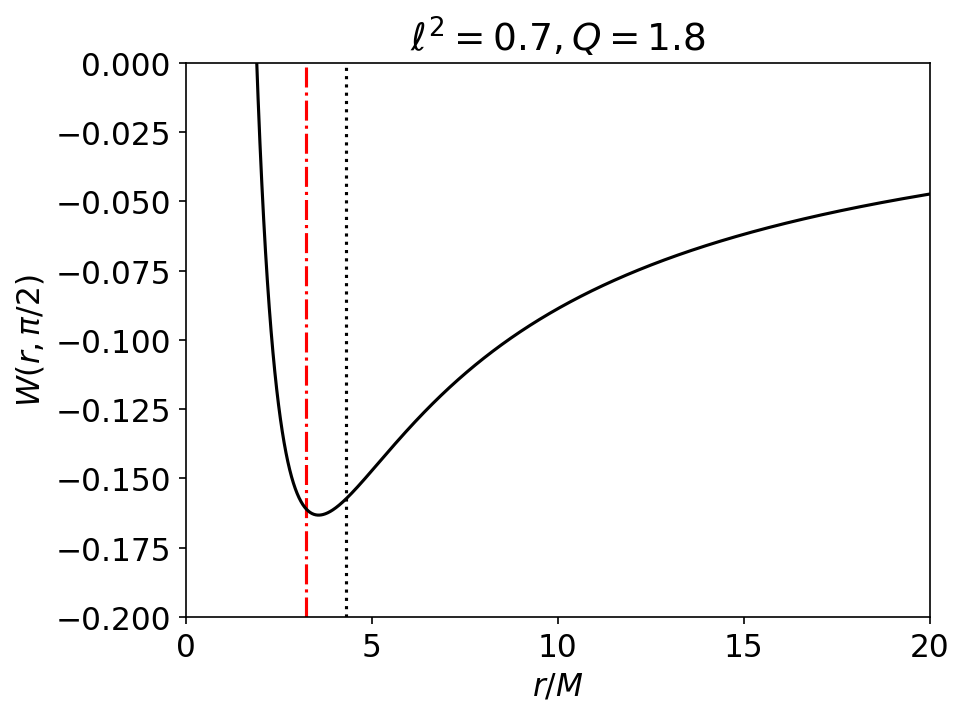}
    \includegraphics[width=1\columnwidth,height=0.65\columnwidth]{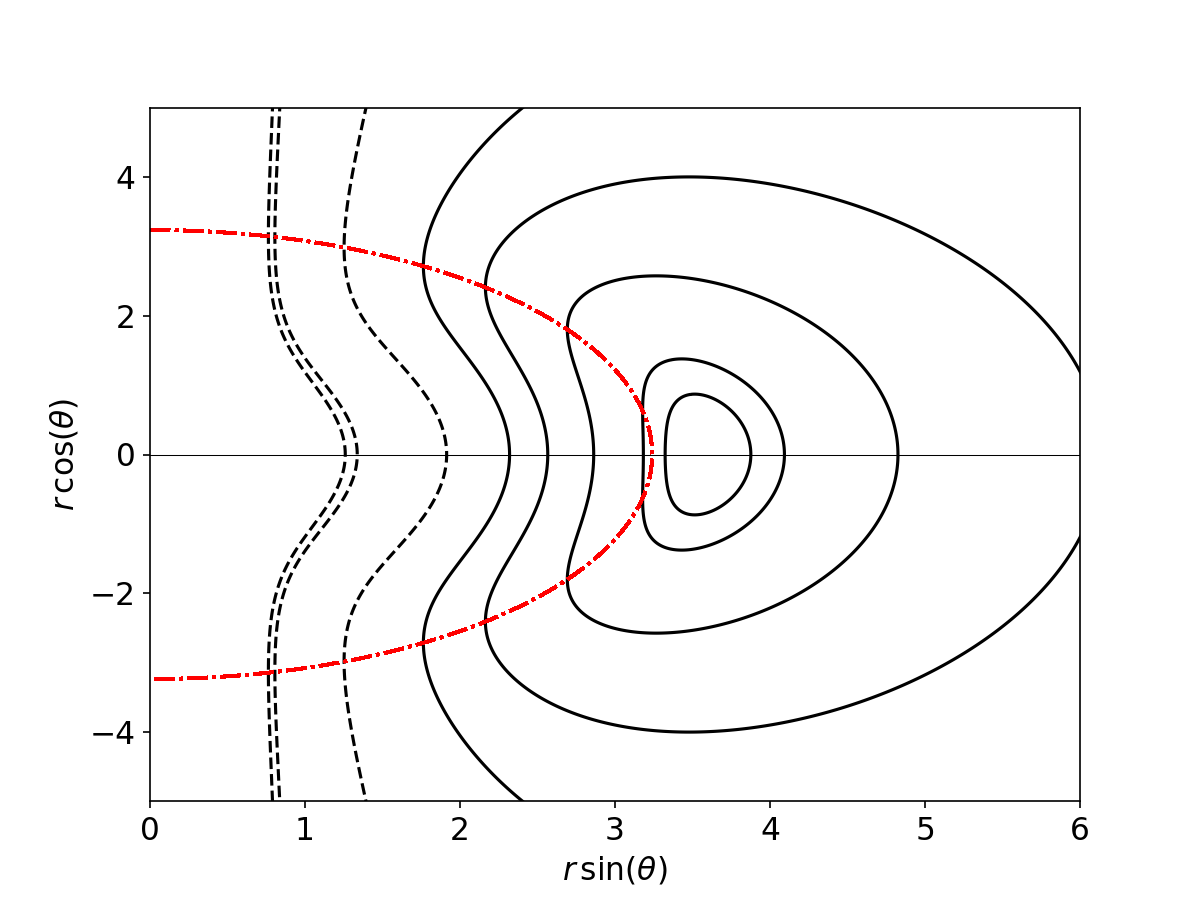}
    \caption{Equilibrium tori around RN naked singularity with $Q/M$=1.8 with $l_0 <l_\mathrm{\Omega max}$.  Location of the zero-gravity sphere is indicated by red dotted-dashed line. {\sl Top:} $W$ as a function of $r/M$ in equatorial plane. The radius of maximum angular frequency, $ r_{\Omega\mathrm{max}}$, is indicated by the black dotted line. {\sl Bottom:} Meridional cross-section of the equipotential surfaces.  Solid closed lines represent toroidal surfaces with negative $W$ value. The black dashed lines represent positive values of W.}
    \label{1b}
\end{figure}
%############################################################################################################################################################################################################################################

For the simplest case of tori with uniform distribution of specific angular momentum, $l_0(r,\theta)= $ constant, $W$ is given by the following equation \citep{2013Rezzolla}:

\begin{equation*}
    W(r,\theta) = \ln |u_t| = \frac{1}{2}\ln \Bigg[\frac{f(r) r^2 \sin^2\theta}{r^2 \sin^2\theta-f(r)l_0^2}\Bigg].
\end{equation*}
At a given point in the $(r,\theta)$ plane, the potential $W$ can be positive or negative indicating open and closed equipotential surfaces. In principle, the fluid can fill any of the closed surfaces, thus a closed equipotential surface can be taken to represent the surface of a stationary configuration of orbiting fluid. The pressure of the toroidal barytrope is zero at the surface, with a non-zero gradient.

For RN naked singularity for constant specific angular momentum $l_0$, the potential $W$ takes the following form:
\begin{align}
    W(r,\theta) = \ln |u_t| =  \frac{1}{2}\ln \Bigg[\frac{(r^2 - {2M}{r} + {Q^2}) \sin^2\theta}{r^2 \sin^2\theta- (r^2 - {2M}{r} + {Q^2})l_0^2/r^2}\Bigg].
\label{surfaces}
\end{align}

%%%%%%%%%%%%%%%%%%%%%%%%%%%%%%%%%%%%%%%%%%%%%%%%
\section*{Toroidal Perfect fluid configurations}
%%%%%%%%%%%%%%%%%%%%%%%%%%%%%%%%%%%%%%%%%%%%%%%%

We construct equilibrium toroidal configurations for four representative values of the angular momentum parameter, $l_0$, for each of two values of the charge to mass ratio (Table~1). These are shown with solid lines in Figs. 4--11. The dashed lines correspond to the surfaces with positive values of $W$ in Eq.~(\ref{surfaces}), so any fluid in their vicinity would be unbound.

The barotropic fluid was assumed to have a uniform distribution of angular momentum. Let us start by discussing the case of $Q/M=1.8$, representative of singularities with no marginally stable orbits.
For low values of specific angular momentum, the circular locus of maximum pressure in the torus (the torus ``center") lies very close to the zero-gravity sphere. While the torus can penetrate the zero-gravity sphere, its gradient of pressure balancing the repulsive gravity, a large part of the fluid must necessarily orbit outside of the zero-gravity sphere. The gravitational repulsion of the naked singularity leads to a very distorted shape of the ``torus," which by analogy to, and in contrast with, the well known black hole ``cusp" solution\footnote{Introduced for the first time in Abramowicz, Jaroszyński \& Sikora 1978.} (known informally as ``Sikora's beak'') could best be described as open ``jaws''  ready to snatch the singularity (Figure~4). For solutions with larger values of angular momentum, these jaws become less prominent and finally disappear as the center of the torus shifts outwards with increasing angular momentum parameters (Figures~5--7). 

For low values of the charge to mass ratio, the presence of marginally stable orbits complicates the solutions. Here, we discuss $Q/M=1.08$, as an example. For low values of angular momentum, the jaws are again apparent inside the zero-gravity sphere (Figure~8). However, for larger values of angular momentum, the torus ends close to the zero-gravity sphere, as though it hit a wall (Figure~9). For larger values of $l_0$ still, two separate toroidal solutions appear (Figure~10). This is related to the two stable orbital solutions for a test particle for a certain range of specific angular momenta (e.g., values of $l_0$ near the maximum of the dashed line in Figure~3). Once angular momentum exceeds the maximum of the $l(r)$ curve, only one stable orbital solution exists, at large radii, and the fluid configuration adopts the shape familiar from studies of compact object gravity in the quasi-Newtonian regime, i.e, far from the source of gravity (Figure~11).
%%%__________________________________________
\begin{table*}
\centering
 \begin{tabular}{||c c c c c ||} 
 \hline
 $Q/M$ & $l_0^2 <<(l_\mathrm{\Omega max})^2$ & $l_0^2 <(l_\mathrm{\Omega max})^2$ & $l_0^2 =(l_\mathrm{\Omega max})^2$ & $l_0^2 >(l_\mathrm{\Omega max})^2$ \\
 \hline\hline
1.08 & 0.05 & 8 &9.73 &16  \\
\hline
1.8 & 0.05 & 0.7& 2.13 & 9 \\
\hline
 \end{tabular}
 \caption{\label{Table 1}List of specific angular momentum values used in our calculations for two different charge to mass ratios.}
\end{table*}
%%%__________________________________________

%############################################################################################################################################################################################################################################
\begin{figure}
    \centering
    \includegraphics[width=1\columnwidth,height=0.65\columnwidth]{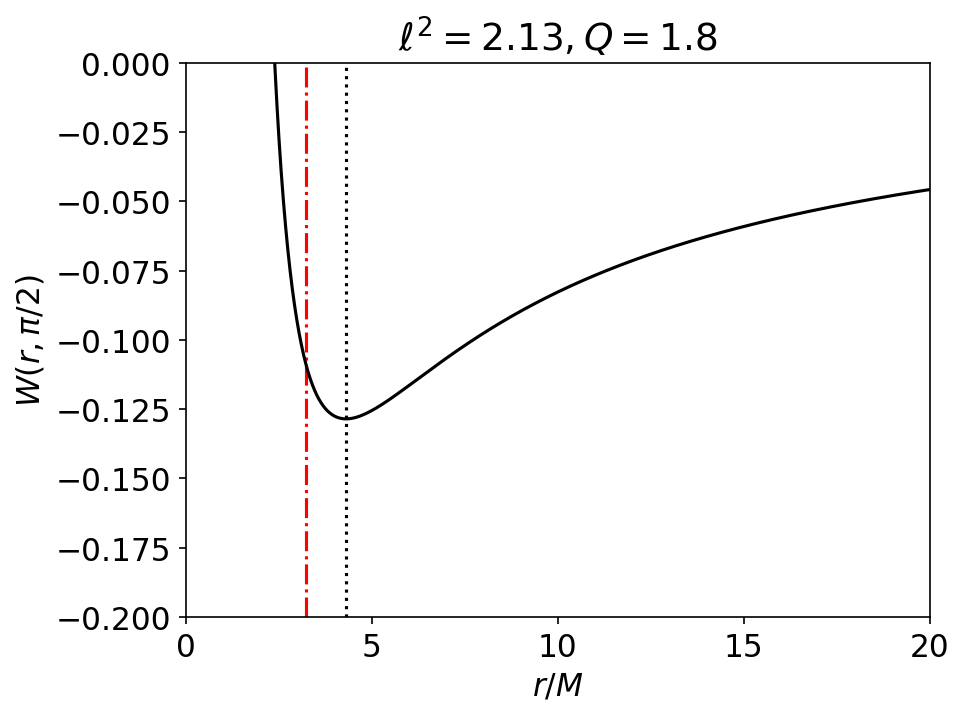}
    \includegraphics[width=1\columnwidth,height=0.65\columnwidth]{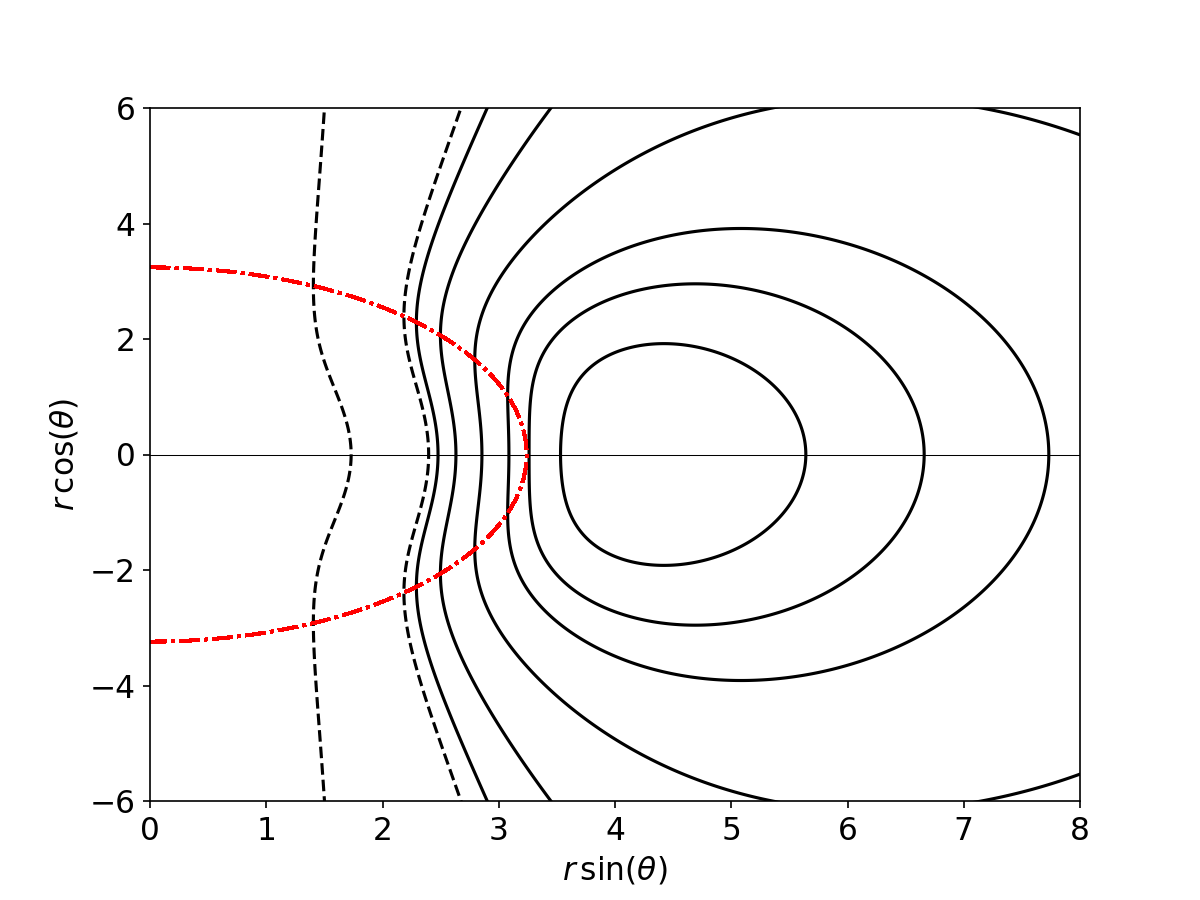}
    \caption{Equilibrium tori around RN naked singularity with $Q/M$=1.8 with $l_0 =l_\mathrm{\Omega max}$.   Location of the zero-gravity sphere is indicated by red dotted-dashed line. {\sl Top:} $W$ as a function of $r/M$ in equatorial plane. The radius of maximum angular frequency, $ r_{\Omega\mathrm{max}}$, is indicated by the black dotted line. {\sl Bottom:} Meridional cross-section of the equipotential surfaces.  Solid closed lines represent toroidal surfaces with negative $W$ value. The black dashed lines represent positive values of W.}
    \label{1c}
\end{figure}
%############################################################################################################################################################################################################################################

%############################################################################################################################################################################################################################################
\begin{figure}
    \centering
    \includegraphics[width=1\columnwidth,height=0.65\columnwidth]{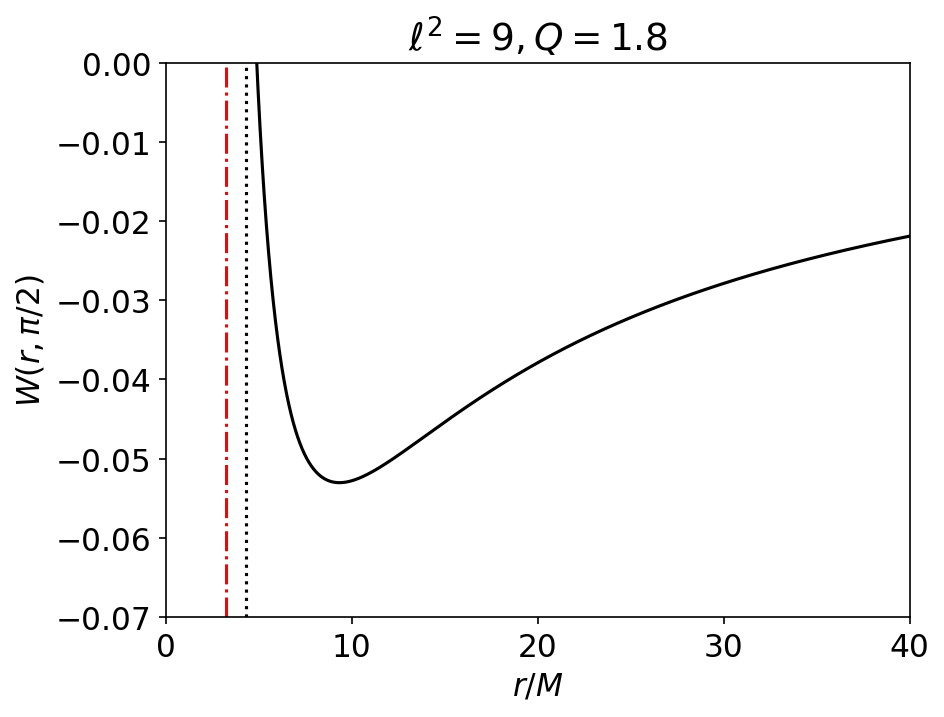}
    \includegraphics[width=1\columnwidth,height=0.65\columnwidth]{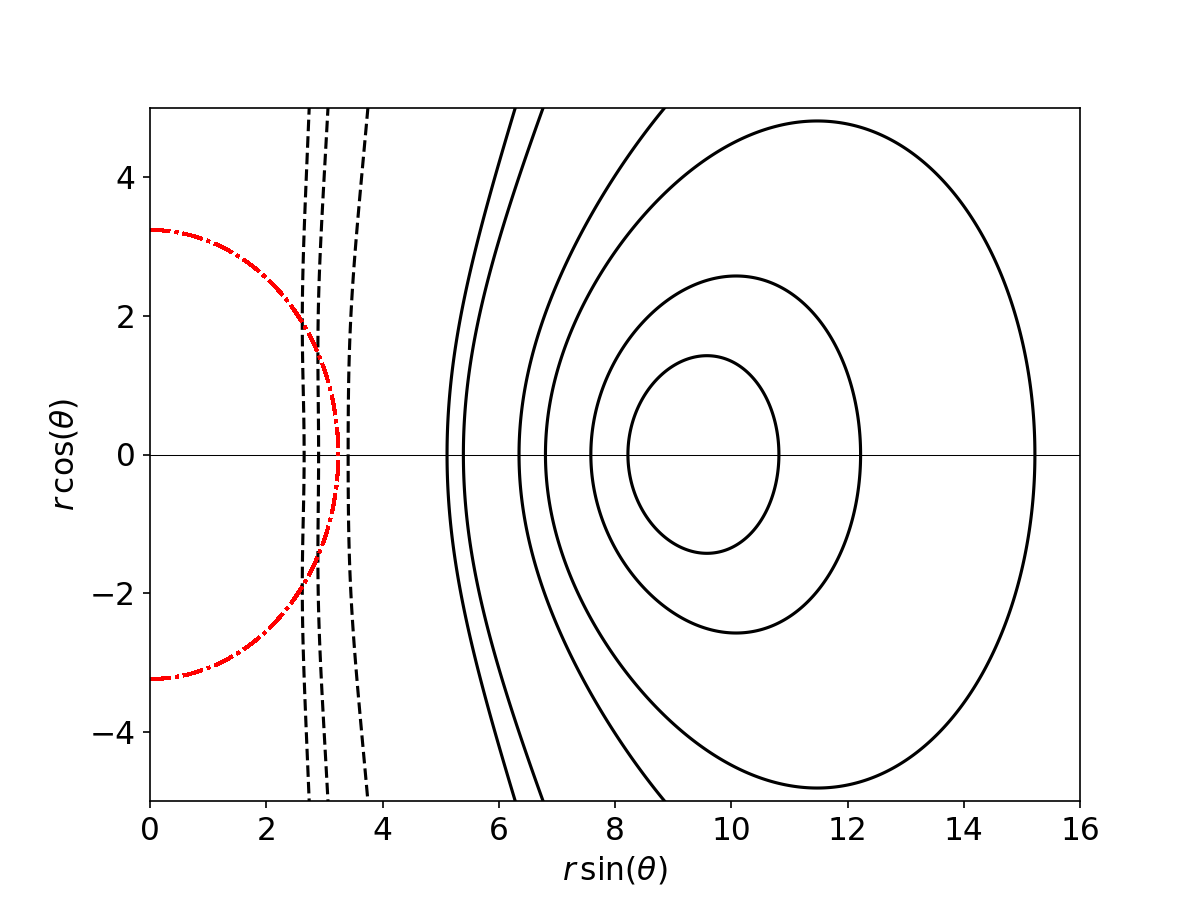}
    \caption{Equilibrium tori around RN naked singularity with $Q/M$=1.8 with $l_0 >l_\mathrm{\Omega max}$.  Location of the zero-gravity sphere is indicated by red dotted-dashed line. {\sl Top:} $W$ as a function of $r/M$ in equatorial plane. The radius of maximum angular frequency, $ r_{\Omega\mathrm{max}}$, is indicated by the black dotted line. {\sl Bottom:} Meridional cross-section of the equipotential surfaces.  Solid closed lines represent toroidal surfaces with negative $W$ value. The black dashed lines represent positive values of W.} 
    \label{1d}
\end{figure}
%############################################################################################################################################################################################################################################

%############################################################################################################################################################################################################################################
\begin{figure}
    \centering
    \includegraphics[width=1\columnwidth,height=0.65\columnwidth]{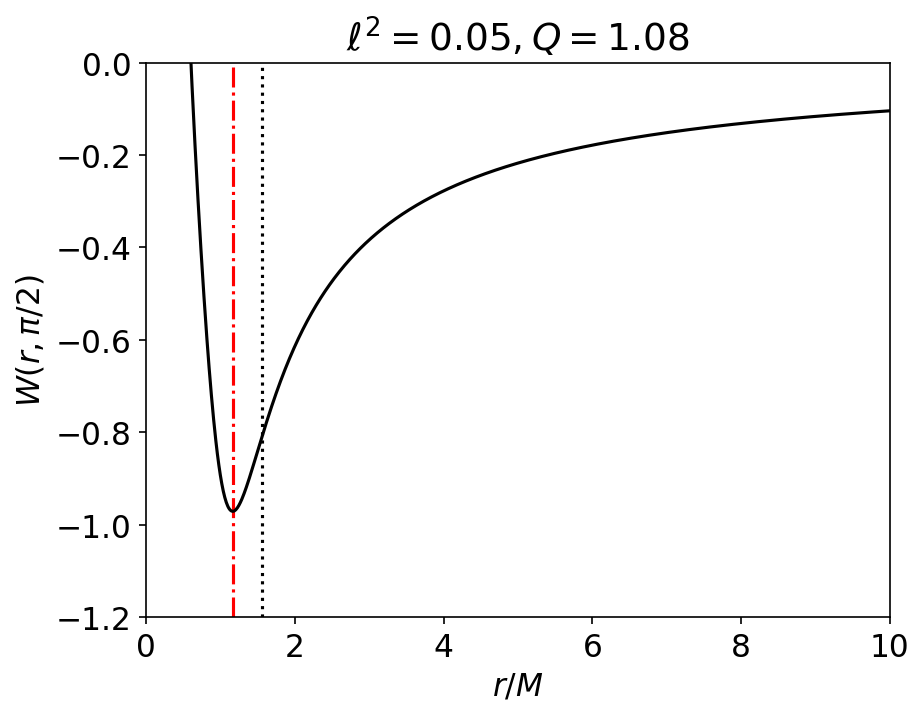}
    \includegraphics[width=1\columnwidth,height=0.65\columnwidth]{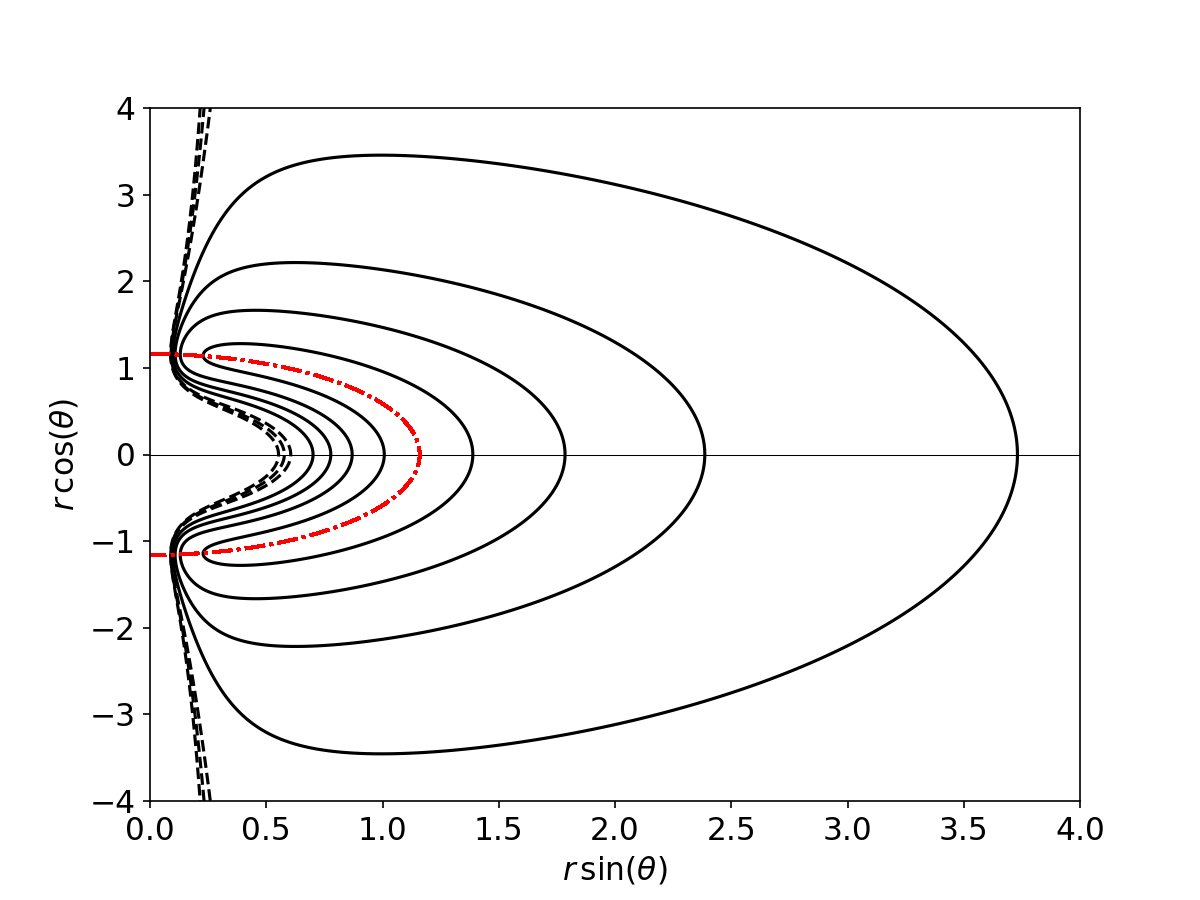}
    \caption{Equilibrium tori around RN naked singularity with $Q/M$=1.8 with $l_0 <l_\mathrm{\Omega max}$.   Location of the zero-gravity sphere is indicated by red dotted-dashed line. {\sl Top:} $W$ as a function of $r/M$ in equatorial plane. The radius of maximum angular frequency, $ r_{\Omega\mathrm{max}}$, is indicated by the black dotted line. {\sl Bottom:} Meridional cross-section of the equipotential surfaces.  Solid closed lines represent toroidal surfaces with negative $W$ value. The black dashed lines represent positive values of W.}
    \label{2a}
\end{figure}
%############################################################################################################################################################################################################################################

%############################################################################################################################################################################################################################################
\begin{figure}
    \centering
    \includegraphics[width=1\columnwidth,height=0.65\columnwidth]{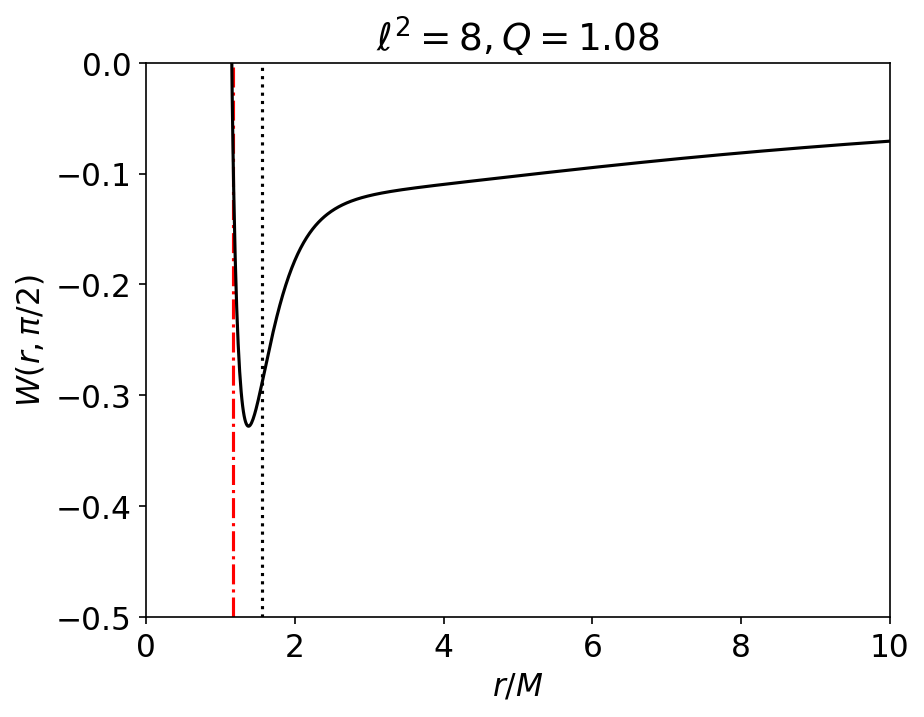}
    \includegraphics[width=1\columnwidth,height=0.65\columnwidth]{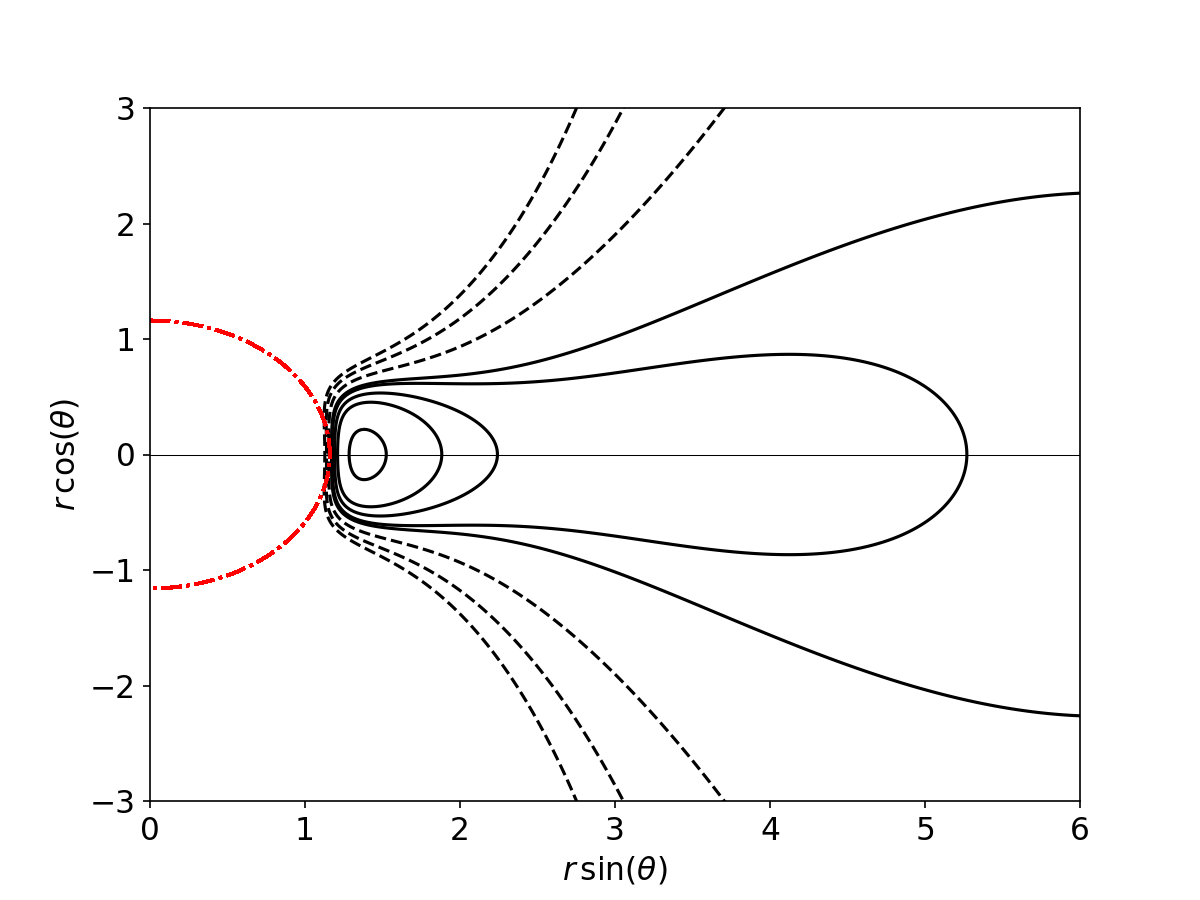}
    \caption{Equilibrium tori around RN naked singularity with $Q/M$=1.8 with $l_0 <l_\mathrm{\Omega max}$.  Location of the zero-gravity sphere is indicated by red dotted-dashed line. {\sl Top:} $W$ as a function of $r/M$ in equatorial plane. The radius of maximum angular frequency, $ r_{\Omega\mathrm{max}}$, is indicated by the black dotted line. {\sl Bottom:} Meridional cross-section of the equipotential surfaces.  Solid closed lines represent toroidal surfaces with negative $W$ value. The black dashed lines represent positive values of W.} 
    \label{2b}
\end{figure}
%############################################################################################################################################################################################################################################

%############################################################################################################################################################################################################################################
\begin{figure}
    \centering
    \includegraphics[width=1\columnwidth,height=0.65\columnwidth]{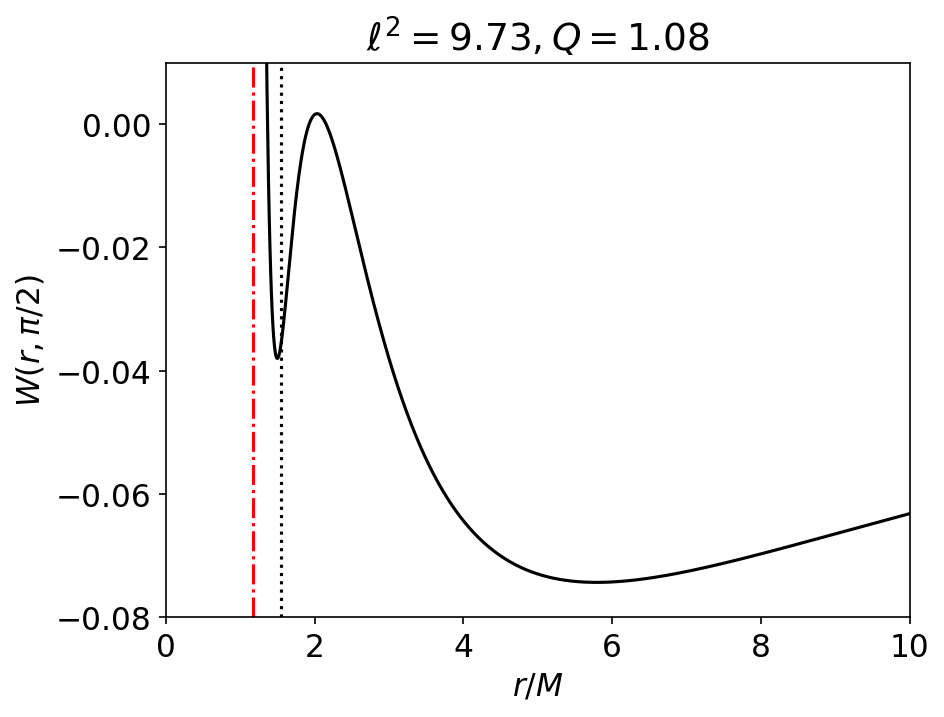}
    \includegraphics[width=1\columnwidth,height=0.65\columnwidth]{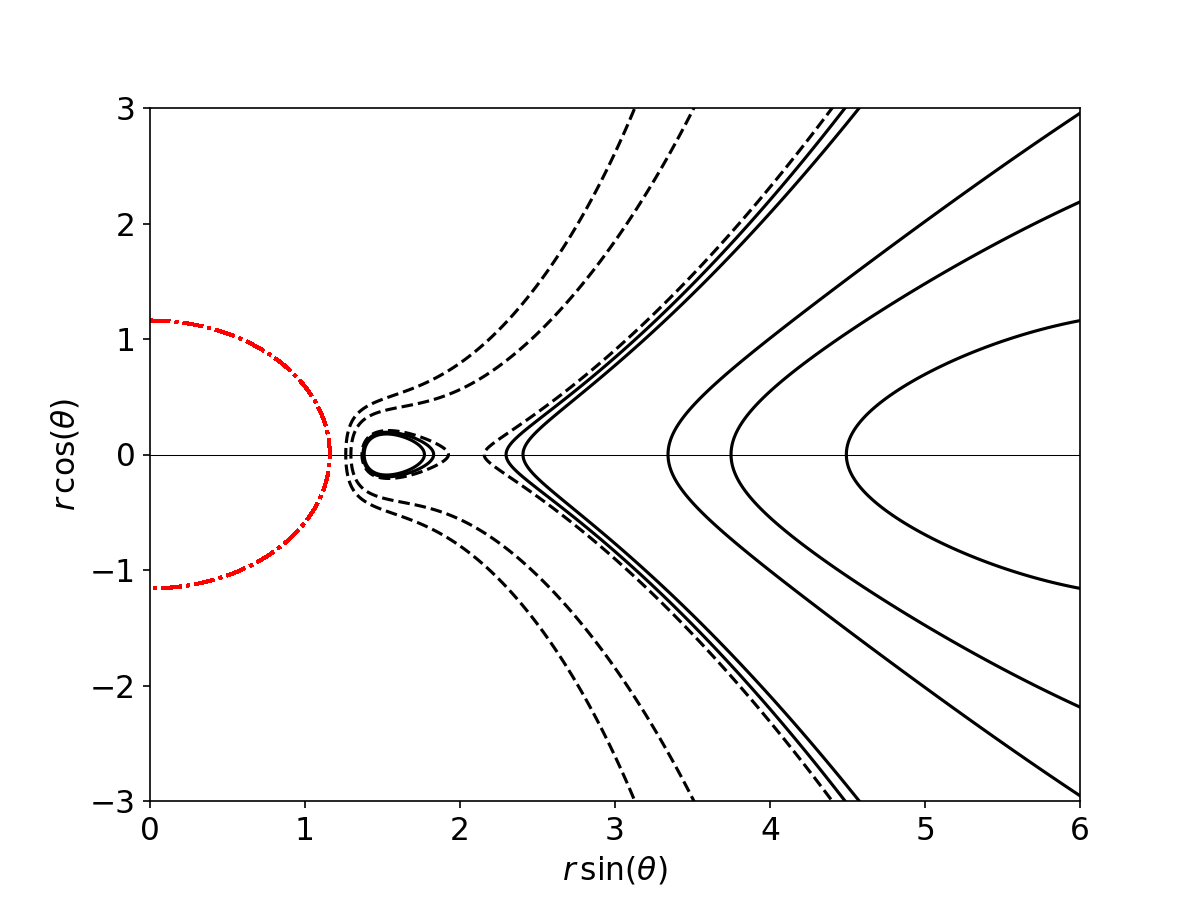}
    \caption{Equilibrium tori around RN naked singularity with $Q/M$=1.8 with $l_0 =l_\mathrm{\Omega max}$.  Location of the zero-gravity sphere is indicated by red dotted-dashed line. {\sl Top:} $W$ as a function of $r/M$ in equatorial plane. The radius of maximum angular frequency, $ r_{\Omega\mathrm{max}}$, is indicated by the black dotted line. {\sl Bottom:} Meridional cross-section of the equipotential surfaces.  Solid closed lines represent toroidal surfaces with negative $W$ value. The black dashed lines represent positive values of W.} 
    \label{2c}
\end{figure}
%############################################################################################################################################################################################################################################

%############################################################################################################################################################################################################################################
\begin{figure}
    \centering
    \includegraphics[width=1\columnwidth,height=0.65\columnwidth]{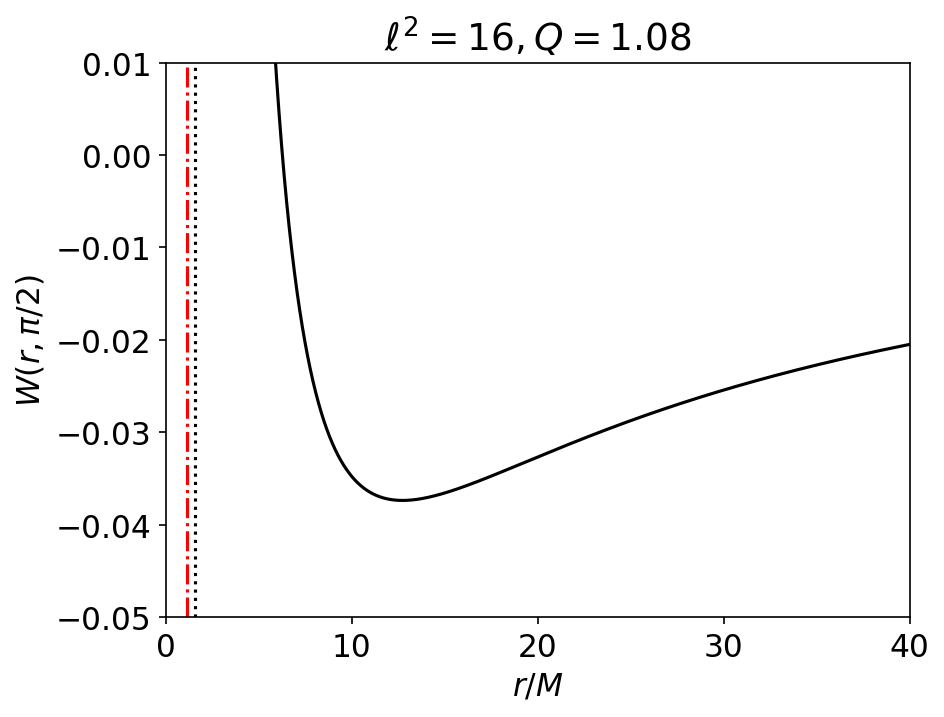}
    \includegraphics[width=1\columnwidth,height=0.65\columnwidth]{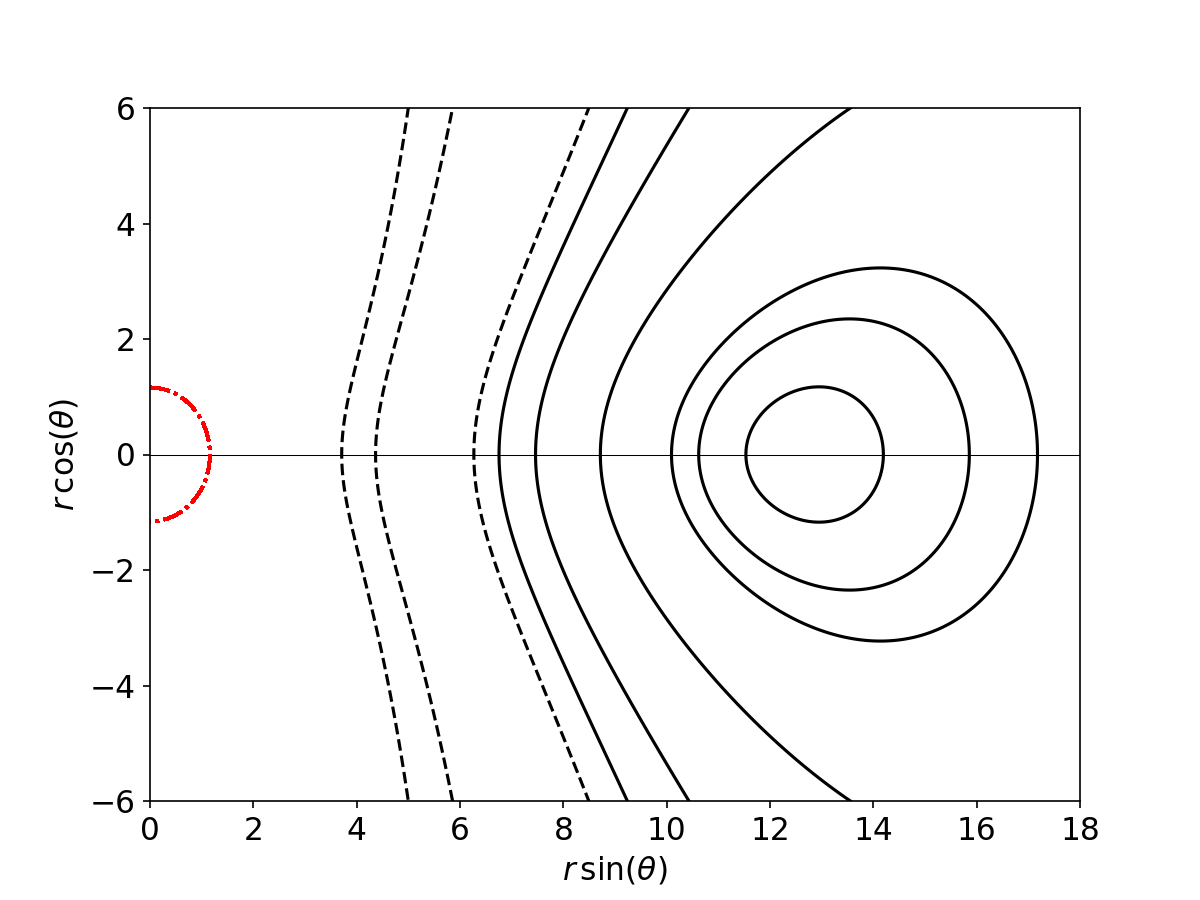}
    \caption{Equilibrium tori around RN naked singularity with $Q/M$=1.8 with $l_0 >l_\mathrm{\Omega max}$.  Location of the zero-gravity sphere is indicated by red dotted-dashed line. {\sl Top:} $W$ as a function of $r/M$ in equatorial plane. The radius of maximum angular frequency, $ r_{\Omega\mathrm{max}}$, is indicated by the black dotted line. {\sl Bottom:} Meridional cross-section of the equipotential surfaces.  Solid closed lines represent toroidal surfaces with negative $W$ value. The black dashed lines represent positive values of W.} 
    \label{2d}
\end{figure}
%############################################################################################################################################################################################################################################
%%%%%%%%%%%%%%%%%%%%%%%%%%%%%%%%%%%%%%%%%%%%%%%%
\section*{Conclusions}
%%%%%%%%%%%%%%%%%%%%%%%%%%%%%%%%%%%%%%%%%%%%%%%%
We have studied the shape of fluid tori orbiting the RN naked singularities. Historically, such studies were performed for black holes to gather insight into accretion disks, in an era before numerical simulations were feasible. This is the first step of a program aiming to retrace this historical development for the case of naked singularities. 

We consider perfect fluid configurations in hydrostatic equilibrium with non-zero\footnote{The zero-angular momentum configurations (levitating atmospheres) are discussed in \cite{2023VK}} uniform angular momentum orbiting Reissner-Nordstr\"om singularities. We find that typically no cusp is formed at the inner edge of toroidal figures of equilibrium around RN singularities. A toroidal solution with a cusp (on the inside of the outer torus, or on the outside of the inner torus) can at most be found in that narrow charge to mass ratio parameter when marginally stable orbits exist (Figure~10). Unlike in black holes, the equipotential surfaces in RN singularities never self-intersect. Thus, unlike in black holes, where accretion from a toroidal ``disk" may proceed onto the compact object through ``Roche-lobe-like" overflow with no angular momentum loss, the RN singularity can only be approached if the accreting fluid loses its excess angular momentum---the zero-gravity sphere is engulfed by the fluid only at low values of specific angular momentum (e.g., Figures 4-6). At higher values of specific angular momentum, only unbound matter may approach the singularity (or even the zero-gravity sphere), but it is likely to outflow to infinity (e.g., Figure 7). 

The positions and shapes of the constant specific angular momentum equilibrium tori, may also be taken to illustrate the general conclusion that for the Reissner-Nordstr\"om naked singularity (and similar space-times) a part of any figure of equilibrium of an orbiting fluid must always lie outside of the zero gravity sphere. This may have applications in interpreting the Event Horizon Telescope (EHT) observations of the center of our Galaxy \citep{2023MV}.
%%%%%%%%%%%%%%%%%%%%%%%%%%%%%%%%%%%%%%%%%%%%%%%%
\ack
%%%%%%%%%%%%%%%%%%%%%%%%%%%%%%%%%%%%%%%%%%%%%%%%
This work was supported in part by the Polish NCN grant 2019/33/B/ST9/01564.
%%%%%%%%%%%%%%%%%%%%%%%%%%%%%%%%%%%%%%%%%%%%%%%%
\bibliography{ragsamp}
%%%%%%%%%%%%%%%%%%%%%%%%%%%%%%%%%%%%%%%%%%%%%%%%
\end{document}